\documentclass[pra,superscriptaddress,twocolumn]{revtex4}
\usepackage{graphicx}
\usepackage{epsfig}
\usepackage{times}
\usepackage{bm}
\usepackage{amsmath}
\begin{document}
\title{High-order above-threshold ionization: the uniform approximation and the
effect of the binding potential}
\author{C. Figueira de Morisson Faria}
\affiliation{Max-Born-Institut, Max-Born-Str. 2a, 12489 Berlin, Germany}
\author{H. Schomerus}
\affiliation{Max-Planck Institut f\"ur Physik komplexer Systeme, N\"othnitzer
Str. 38, 01187 Dresden, Germany}
\author{W. Becker}
\altaffiliation[also at ]{Center for Advanced Studies, Department of Physics
and Astronomy, University of New Mexico, Albuquerque, NM 87131}
\affiliation{Max-Born-Institut, Max-Born-Str. 2a, 12489 Berlin, Germany}

\date{\today}
\begin{abstract}
A versatile semiclassical approximation for intense laser-atom processes is
presented.
 This uniform
approximation is no more complicated than the frequently-used multi-dimensional saddle-point
approximation and far superior, since it    
applies for all energies, both close to as well as
away from classical cutoffs. In the latter case, it reduces to the standard
saddle-point approximation. The uniform approximation agrees accurately with numerical evaluations for potentials, for which these are feasible, and constitutes a practicable method of 
calculation in general. The method is applied to the calculation of  high-order above-threshold
ionization spectra with various binding potentials: Coulomb, Yukawa, and shell
potentials which may model C$_{60}$ molecules or clusters. 
The shell potentials 
generate high-order ATI spectra that are more structured and may 
feature an apparently higher cutoff.
\end{abstract}

\maketitle

\newcommand{\vp}{\mathbf{p}}
\newcommand{\vk}{\mathbf{k}}
\newcommand{\vA}{\mathbf{A}}
\newcommand{\vecr}{\mathbf{r}}

\section{Introduction}

Sufficiently intense laser fields ionize atoms or molecules by the
quantum-mechanical process of tunneling \cite{DK}. Both the tunneling process
and the ensuing motion of the electron in the continuum are well accessible to
semiclassical methods. Tunneling generates a wave packet whose center follows a
classical trajectory while the wave packet is spreading.
It may or may not
return to within the range of the ionic binding potential. If it does, the
well-known recollision-induced processes, such as high-order harmonic
generation (HHG) or high-order above-threshold ionization (ATI), take
place \cite{corkum93,MA}.

In the tunneling regime, the quantum-mechanical transition amplitude can be
analyzed, computed, and interpreted via the saddle-point approximation 
\cite{lew94,lew95}. Typically, the transition amplitude is represented by a 
multi-dimensional integral over the time $t'$ at which the electron enters the
continuum by tunneling, the later time $t$ at which it revisits the ion, and 
one or all components of the drift momentum $\mathbf{k}$ along its orbit in
between those two times. For a specified final state, e.g., for given final
momentum of the electron after the recollision or for given frequency of
high-order harmonic emission, the saddle-point approximation selects those
particular ``quantum orbits'' that contribute to this final state. These orbits
are characterized by particular values of the parameters $t,t'$ and
$\mathbf{k}$, which are complex numbers because of the tunneling nature of
these orbits. For a specified final state, there are, in general, several 
contributing quantum orbits. Their contributions have to be added coherently,
and this yields an interference
pattern, which
may appear very intricate, even though
its physical origin is simple \cite{optcomm,science}.

Within the context of atoms in strong fields, the contributing quantum 
orbits typically come in pairs.
This may be best known from the Lewenstein model of HHG: 
For specified harmonic order within the
``plateau'', there are  two  quantum orbits whose contributions dominate the
harmonic yield, the ``long orbit'' and the ``short orbit''. An electron on the
long orbit starts earlier (by ionization) and returns later (for recombination)
than an electron on the short orbit \cite{lew94}. This is a very general
feature of intense-laser--atom processes and holds also for the more
complicated orbits, which bypass the ion once or several times before the
recombination process takes place \cite{optcomm}. For fixed laser intensity,
the maximal HHG frequency or the maximal energy of an ATI electron obey 
classical limits \cite{corkum93,kulander}, which are 
related to the maximal kinetic energy of the electron returning to the ion.
For parameters approaching such classical limits, the
two quantum orbits become more and more identical. If it were not for the fact
that their parameters are complex, reflecting the birth of the electron by
tunneling, the two orbits of a pair would coalesce at the classical cutoff
\cite{optcomm}.

The near coalescence of the orbits of a pair near a cutoff constitutes two
problems for the saddle-point approximation: (i) treating the two saddle points
 as independent becomes an increasingly inaccurate approximation if they
approach each other closely \cite{goresl,GP00}; should they actually merge into one,
the standard saddle-point approximation diverges and hence is completely
inapplicable; (ii) beyond the cutoff, in the classically forbidden regime, both
(complex) saddle points continue to exist as formal solutions of  
the saddle-point
conditions.  One, however, has to be dropped from the transition
amplitude, which is frequently, but not always, indicated
by an exponentially exploding contribution. A
rigorous analysis establishes that, actually, it is not possible to deform
the contour of integration through such a saddle point by the method of
steepest descent. In the framework of the theory of asymptotic expansions,
the global bifurcation of
the steepest-descent contour from two visited saddles to a single
visited saddle is known as the Stokes transition \cite{BH,Berry}.
In previous work, these problems were not treated in a systematic fashion. In
this paper, we invoke a specific uniform approximation to solve both
problems \cite{Henning1}.
This turns out to be no more complicated than the standard procedure
of treating the two saddle points as independent, because it uses exactly
the same information as the standard procedure, namely, the values of the 
action and its
second derivative at the saddle points. Problem (i) is solved because the
uniform approximation regularizes the saddle-point integrals close to the
classical cutoff, while it reduces to the saddle-point approximation far away
from the cutoff.
Problem (ii) is solved by imposing the simple
requirement of continuity on the transition amplitude, which
automatically selects the
appropriate branch of the multivalued solution that does not contain the
contribution of the unphysical saddle point beyond the classical cutoff.
For a zero-range binding potential, the benefit of the
saddle-point approximation lies in the insight gained by the
introduction of a few quantum orbits, which allow one to visualize the physical
mechanism behind recollision-induced processes.  For the mere purpose of
 computation,  the transition amplitude can be
calculated as well, if not more easily, via a simple quadrature. 
We will use the zero-range potential as a test
case, and find excellent agreement for the uniform approximation, even
where the usual saddle-point approximation fails.

The zero-range potential is a valid model for the description of a negatively
charged ion in an intense laser field \cite{0range,manakov,helm}. To what extent it can also be employed to model an atom in an
intense laser field or, in other words, just how important the long range of
the Coulomb potential is in this situation, has been the object of some debate.
Surprisingly, it has turned out that at least for the qualitative explanation
of most intense-field effects the Coulomb tail is not instrumental \cite{lew94,lew95,aamop02}.
Still more
surprisingly, even the subtle quantum-mechanical enhancements of the ATI
plateau at certain sharply defined intensities \cite{ccres} are not specific to the Coulomb
potential. In fact, a zero-range potential yields virtually the same
enhancements, though at slightly different intensities \cite{0rangeres}.

From this point of view, being able to compare ATI spectra from zero-range 
and non-zero-range potentials is important. However, for non-zero-range 
potentials, a direct
computation of the  transition amplitude requires one to carry out a
cumbersome multidimensional integral, and the uniform saddle-point 
approximation is the only viable approach.

The purpose of this paper then is twofold. First, we determine the
specific uniform
approximation that applies to the pairs of quantum orbits that appear in
laser-induced rescattering processes. Second, we use this uniform approximation
to investigate the influence of the form of the binding potential on 
ATI.

The plan of the paper is as follows: In Sec.\ \ref{sec:keldysh},
we summarize the improved Keldysh approximation for the transition amplitude. 
In Sec.\ \ref{sec:saddles}, we
discuss the saddle points that feature in the saddle-point approximation
as well as in the uniform approximation, and review the saddle-point
approximation as well as its problems close to classical cutoffs.
In Sec.\ \ref{uniform}, we determine the uniform
approximation that overcomes these problems
and describe its conceptual relation to the saddle-point
approximation.
In Sec.\ \ref{sec:compare} we compare the ATI spectra obtained by these
approximations to the numerical results for the zero-range binding
potential.
The uniform approximation is then used
 in Sec.\ \ref{sec:potential} to address the effect of a general (non-zero-range) binding potential on the ATI spectrum, using 
 Coulomb, Yukawa, and shell potentials as examples.
A summary of the results and conclusions can be found
in Sec.\ \ref{sec:conclusions}.

We use atomic units (a.u.) throughout this paper. 
\section{Transition amplitude for rescattering processes}
\label{sec:keldysh}

Strong-field phenomena, such as above-threshold ionization (ATI), 
are successfully described by
transition amplitudes derived within a framework known as the strong-field
approximation. This approximation neglects the
binding potential in the propagation of the electron in the continuum, and
the laser field when the electron is bound, which corresponds to
treating the process of rescattering in
the first-order Born approximation on the background of the laser field.
(The first-order Born approximation yields the exact differential cross
section in the absence of the field both for the Coulomb potential as well as
for the zero-range potential.)
The ATI transition amplitude for
the direct electrons -- electrons that leave the vicinity of the ion right
after they have tunneled into the continuum --  is  the well-known
Keldysh-Faisal-Reiss (KFR) amplitude \cite{KFR}
\begin{equation}
M_{\mathrm{dir}}=-i \int_{-\infty }^{\infty}dt'\,
 \langle \psi _{\vp}^{(V)}(t') |V| \psi_{0}(t')\rangle .  \label{direct}
\end{equation}
The generalized transition amplitude,  which includes one single act of rescattering, is
given by \cite{LKKB}
\begin{equation}
M_{\mathrm{resc}}=-\int_{-\infty }^{\infty }dt\int_{-\infty }^{t}dt^{\prime
}\, \langle \psi _{\vp}^{(V)}(t)| VU^{(V)}(t,t^{\prime })V | \psi
_{0}(t^{\prime })\rangle .  \label{rescatt}
\end{equation}
In both equations,
$V$ denotes the atomic binding potential,
the final state is the Volkov state describing a charged particle 
with asymptotic momentum $\vp$ in the 
presence of a field with vector potential $\vA(t)$,
\begin{equation}
\langle \vecr|\psi _{\vp}^{(V)}(t)\rangle = \exp\left(-\frac{i}{2}\int_t^\infty d\tau\,
[\vp+\vA(\tau)]^2 \right)
e^{i[\vp+\vA(t)]\cdot \vecr}, \label{volkov}
\end{equation}
and $U^{(V)}(t,t^{\prime })$
is the Volkov
time-evolution operator, which describes the evolution of the electron in the presence of only the laser field. In Eq.~(\ref{direct}), the electron, initially in the
ground state $| \psi _{0}(t')\rangle$, is ionized into its final state  at the
time $t'$. In Eq.~(\ref{rescatt}), an additional rescattering off the binding
potential
at the time $t$ is accounted for. The amplitude (\ref{rescatt}) 
incorporates the amplitude (\ref{direct}) for direct ionization in the limit
where $t' \rightarrow t$. Hence, the two amplitudes must not be added
\cite{LKKB}. The amplitude (\ref{rescatt}) or closely related versions thereof
have been  used by several authors \cite{milos,goresl,GP00,goresl2}.

If we insert the expansion of the Volkov propagator in terms of Volkov states,
\begin{equation}
U^{(V)}(t,t^{\prime }) = \int d^3\vk\, |\psi _{\vk}^{(V)}(t)\rangle \langle \psi
_{\vk}^{(V)}(t')|, \label{volkovexpand}
\end{equation}
 into Eqs.\
(\ref{direct}) and (\ref{rescatt}),
the transition amplitudes can be rewritten as
\begin{equation}
M_{\mathrm{dir}}=-i\int_{-\infty }^{\infty}dt'\, \exp [iS_{\vp}(t')]\, V_{\vp 0},  \label{pdirect}
\end{equation}
and
\begin{equation}
M_{\mathrm{resc}}=-\int_{-\infty }^{\infty }dt\int_{-\infty
}^{t}dt^{\prime }\int d^{3}\vk\, e^{iS_{\vp}(t,t^{\prime },\vk)} V_{\vp \vk}V_{\vk
0},
\label{prescatt}
\end{equation}
where the corresponding actions are given by
\begin{equation}
S_{\vp}(t')=-\frac{1}{2}\int_{t^{\prime }}^{\infty }d\tau \,\left[
\vp+\vA(\tau )\right] ^{2}+ |E_{0}|t \label{daction}
\end{equation}
and
\begin{eqnarray}
S_{\vp}(t,t',\vk) &=&-\frac{1}{2}\int_{t}^{\infty }d\tau \,\left[ \vp+\vA(\tau
)\right] ^{2} \nonumber\\
&-&\frac{1}{2}\int_{t^{\prime }}^{t}d\tau\,
\left[ \vk+\vA(\tau )\right] ^{2}+ |E_{0}|t'. \label{raction}
\end{eqnarray}
The quantity $|E_{0}|$  denotes the ionization potential of the atom. 
In this paper, we address the case of a linearly polarized monochromatic field,
\begin{equation}
\vA(t)=A_0\mathbf{e}_x \cos \omega t,
\label{eq:linpol}
\end{equation}
with the ponderomotive energy $U_P=\langle \vA^2(t)\rangle _t/2=A_0^2/4$.

The representations (\ref{pdirect}) and (\ref{prescatt})
are particularly useful if the form factors
\begin{eqnarray}
V_{\vp \vk} &=&\left\langle \vp+\vA(t)\right| V\left| \vk+\vA(t)\right\rangle
\nonumber \\ &=&\frac{1}{(2\pi )^{3}}\int d^{3}\vecr\,\exp [-i(\vp-\vk)\cdot
\vecr]V(\vecr) \label{ffr}
\end{eqnarray}
and
\begin{eqnarray}
V_{\vk 0}
&=&
\left\langle \vk+\vA(t^{\prime })\right| V\left|
0\right\rangle\nonumber \\
&=&
\frac{1}{(2\pi )^{3/2}}\int d^{3}\vecr\,\exp [-i(\vk+\vA%
(t^{\prime }))\cdot \vecr]V(\vecr) \psi_0(\vecr) 
\nonumber \\ \label{ffi}
\end{eqnarray}
can be calculated in analytical form.
Within the strong-field approximation, the influence of
the binding potential is entirely
contained in these two matrix elements. For a zero-range potential, the form
factors are constants. In this case, the five-dimensional integral
(\ref{prescatt}) can be reduced to a one-dimensional integral over a series of
Bessel functions, which can be readily computed numerically \cite{LKKB,JPB}. In
Sec.\ \ref{sec:compare},
we will refer to the outcome of this procedure as the ``exact result''. 
In general, however, a correspondingly ``exact'' evaluation of the matrix element (\ref{rescatt}) has to deal with a multidimensional integral.

\section{Saddle-point analysis}
\label{sec:saddles}

For sufficiently high intensity
of the laser field,
corresponding to 
small Keldysh parameter $\gamma=\sqrt{|E_0|/2U_P}$,
ionization can be envisioned to proceed via the quasistatic process of tunneling  \cite{footnkeldysh}.
The transition amplitudes (\ref{pdirect}) and (\ref{prescatt}) are then 
conveniently computed via the method of steepest descent.
Both the standard saddle-point approximation 
as well as the uniform
approximation 
rest on this method, which 
approximates the entire integral by
the contributions from the vicinity of those points on the integration contour
where the action is stationary, i.e., where the partial derivatives of
the action with respect to
the integration variables vanish.
These points correspond to maxima of the
integrand after a deformation of the original integration manifold,
which is constructed such that the integrand decreases
roughly like a Gaussian when one moves away from the vicinity of
the saddles \cite{BH}.

In the current section, we first write down the equations that determine
 the saddle points,
then describe the general procedure of identifying the \textit{relevant} 
saddles,
and finally discuss the saddle-point approximation. All these items are
prerequisites for the discussion of the uniform approximation 
in Sec.\ \ref{uniform}.

\subsection{Saddle-point equations}

For the rescattering amplitude (\ref{prescatt}), the
saddle-point equations are
\begin{eqnarray}
\left[ \vk+\vA(t^{\prime })\right] ^{2}&=&-2|E_{0}|,  \label{saddle1}\\
\left[ \vp+\vA(t)\right] ^{2}&=&\left[ \vk+\vA(t)\right] ^{2},
\label{saddle2}\\
\int_{t^{\prime }}^{t}d\tau \left[ \vk+\vA(\tau )\right] &=& 0.
\label{saddle3}
\end{eqnarray}
Their solutions determine the ionization time $t'$, the rescattering time $t$, and
the drift momentum $\vk$ of the electronic orbit in between those two times,
such that the electron acquires the asymptotic momentum $\vp$.
Equations~(\ref{saddle1}) and (\ref{saddle2}) are related to energy
conservation at the ionization time and
the rescattering time, respectively, and Eq.~(\ref{saddle3}) determines
the intermediate electron momentum.
For the direct
amplitude (\ref{pdirect}), only the ionization time $t'$ need be determined,
and the resulting equation is like Eq.~(\ref{saddle1}) with $\vk$ replaced by
the asymptotic momentum $\vp$.

Evidently, Eq.~(\ref{saddle1}) has no real
solutions $t'$ as long as $E_0 \neq 0$,
and in consequence $t,t^{\prime }$ and $\vk$ are complex.
Physically, the fact that $t'$ is complex means that ionization
takes place through a tunneling process. 
The solutions $(t,t^{\prime })$ of the saddle-point equations for the 
linearly polarized monochromatic field (\ref{eq:linpol}) 
have been computed in Ref.~\cite{optcomm}. They only depend on
the ionization energy $E_0$ and the photoelectron momentum $\vp$,
but not on the shape of the binding potential,
which
enters the transition amplitude only via the form factors 
(\ref{ffr}) and (\ref{ffi}).

A very important feature of the solutions is that they come in pairs. Let us
denote the ``travel time'' by $\tau \equiv t - t'$. Then, for given asymptotic
momentum $\vp$ and for the $n$th travel-time  time interval  $nT/2 \le
\textrm{Re}\, \tau \leq (n+1)T/2 \ (n=1,2,\dots)$, there are two solutions
having slightly different travel times.
The parameters of two typical pairs of quantum orbits are displayed in
Fig.\ \ref{figsaddles}.

\subsection{Classical cutoffs and Stokes transitions}

The original contour of integration in the amplitudes (\ref{pdirect}) or
(\ref{prescatt}) is along the real axes, while the solutions of the
saddle-point equations (\ref{saddle1})--(\ref{saddle3}) are located off the
real axes in the complex plane.
A central question in the method of steepest descent then is, which of the
various saddle points are visited by the steepest-descent integration manifold.
We shall call those the \textit{relevant} saddle points.
The steepest-descent manifold consists of pieces 
with a constant
real part of the action.
These pieces are glued together
at zeros of the integrand, at which the phase of the action is not
well defined.
Usually, each piece visits only a single saddle point, which also
determines the constant
real part of the action.
Only such pieces that are needed to connect the integration
boundaries give contributions to the transition amplitude.
The number of these pieces can change
in a so-called Stokes transition, when two pieces merge at a certain
value of a parameter (here we consider the photoelectron momentum
$\vp$).
On either side of the Stokes transition, the manifolds 
of the saddles of interest
are glued together in different ways: on one side,  
 both pieces are needed to connect the integration
boundaries (plus, possibly, other pieces related to different 
pairs of saddle points), while only one of the
pieces is needed on the other.
Note that in the latter case, too, 
there are still two solutions of the saddle-point equations,
but only one of them is visited by the steepest-descent deformation of
the original integration manifold \cite{kopoldthesis}.

Merging of steepest-descent manifolds requires that the real parts of the
actions of two quantum orbits become identical at a specific value of
$\vp$,
\begin{equation}
\mathrm{Re}\,S_\vp(t_{i},t^{\prime}_{i},\vk_{i})=
\mathrm{Re}\,S_\vp(t_{j},t^{\prime}_{j},\vk_{j})
,
\label{stokesline}
\end{equation}
where $i$ and $j$ denote the saddle points of the given pair, and the times $t_s$
and $t'_s\ (s=i,j)$ depend on $\vp$.
It follows from the  physical mechanism behind high-order ATI that
both saddles of each pair are relevant provided the asymptotic
momentum is classically accessible. For the pair of orbits
having the shortest travel times $(n=1)$, this is the case if
$\vp^2/2 \le 10.007 U_P$
\cite{PBW95}. The other pairs of orbits have smaller cutoff
energies.

The relevant saddle beyond the classical cutoff is the one that has the smaller
imaginary part
of the action at the Stokes transition \cite{footnote27}.
In the
following we reserve the index $i$  for this saddle.
Saddle $j$ only maintains a residual contribution
to the transition amplitude  after the Stokes transition,
until it becomes completely irrelevant in the
so-called anti-Stokes transition
\begin{equation}
\mathrm{Im}\,S_\vp(t_i,t^{\prime}_i,\vk_{i})=
\mathrm{Im}\,S_\vp(t_j,t^{\prime}_j,\vk_{j}).
\label{eq:antistokes}
\end{equation}
The anti-Stokes transition
coincides with the Stokes transition if both saddles actually
coalesce. Otherwise, it frequently
occurs very shortly after the Stokes transition.

Exactly how the transition amplitude
behaves close to the classical cutoff can only be described when the 
interplay of both saddles is taken into account
in a systematic way, which is achieved by the
uniform approximation. Before we turn to this approximation, we now
discuss the standard saddle-point approximation.

\subsection{Saddle-point approximation}
\label{sec:spa}

Within the saddle-point approximation, the amplitudes (\ref{pdirect}) and
(\ref{prescatt}) are approximated by
\begin{equation}
M_{\mathrm{dir}}^{(\mathrm{SPA})}=\sum_{s}\sqrt{\frac{2\pi i}{\partial
^{2}S_{\vp}/\partial t_{s}^{2}}}V_{\vp 0}\exp [iS_{\vp}(t_{s})]
\label{saddirect}
\end{equation}
and
\begin{subequations}
\label{sadresc}
\begin{eqnarray}
M_{\mathrm{resc}}^{(\mathrm{SPA})}&=&\sum_{s} A_s\exp(iS_s)
,
\label{sadresca}
\\
S_s&=&S_{\vp}(t_s,t'_s,\vk_s)
,
\\
A_s&=&(2\pi i)^{5/2}
\frac{V_{\vp \vk_s}V_{\vk_s 0}
}{\sqrt{\det
S''_\vp(t,t',\vk)|_s}}  
,
\label{sadrescc}
\end{eqnarray}
\end{subequations}
respectively, where the index $s$ runs over the relevant saddle points, and
$S''_\vp(t,t',\vk)|_s$ is the five-dimensional matrix of the second derivatives
of the action (\ref{raction}) evaluated at the  solutions of the saddle-point
equations (\ref{saddle1})-(\ref{saddle3}).

In explicit calculations, we will proceed slightly differently: First, we
employ the saddle-point approximation to evaluate the three-dimensional
integral over the intermediate momentum $\vk$ in Eq.~(\ref{prescatt}), which
enters the action (\ref{raction}) only quadratically. This results in
\begin{equation}
M_{\mathrm{resc}}=-\int_{-\infty }^{\infty }dt\int_{-\infty}^{t}
dt^{\prime } e^{iS_{\vp}(t,t^\prime)} V_{\vp \vk(t,t')}V_{\vk(t,t') 0},
 \label{prescattmod}
\end{equation}
where
\begin{equation}
\vk(t,t')=-\frac{1}{t-t'}\int^t_{t'}d\tau \vA(\tau) \label{kstat}
\end{equation}
and $S_{\vp}(t,t^\prime)\equiv S_\vp (t,t',\vk(t,t'))$. Then, we again make use
of the saddle-point approximation to compute the two-dimensional integral over
$t$ and $t'$ in Eq.~(\ref{prescattmod}), which again results
in the amplitude (\ref{sadresc}), where the actions and amplitudes
are now computed by
\begin{subequations}
\label{sadresc2}
\begin{eqnarray}
S_s&=&S_{\vp}(t_s,t'_s)
,
\\
A_s&=&
(2\pi i)^{5/2}
\frac{V_{\vp \vk(t_s,t'_s)}V_{\vk(t_s,t'_s) 0}
}{
\sqrt{(t_s'-t_s)^3\det
S''_\vp(t,t')|_s}}
\label{sadresc2c}
.
\end{eqnarray}
\end{subequations}
The corresponding saddle-point
equations are Eqs. (\ref{saddle1}) and (\ref{saddle2}) with $\vk$ replaced by
$\vk(t,t')$.
Note that the values $S_s$, $A_s$ of each saddle point are not changed,
they are just obtained from a different set of relations
in this more practical procedure.

\begin{figure*}
\epsfig{file=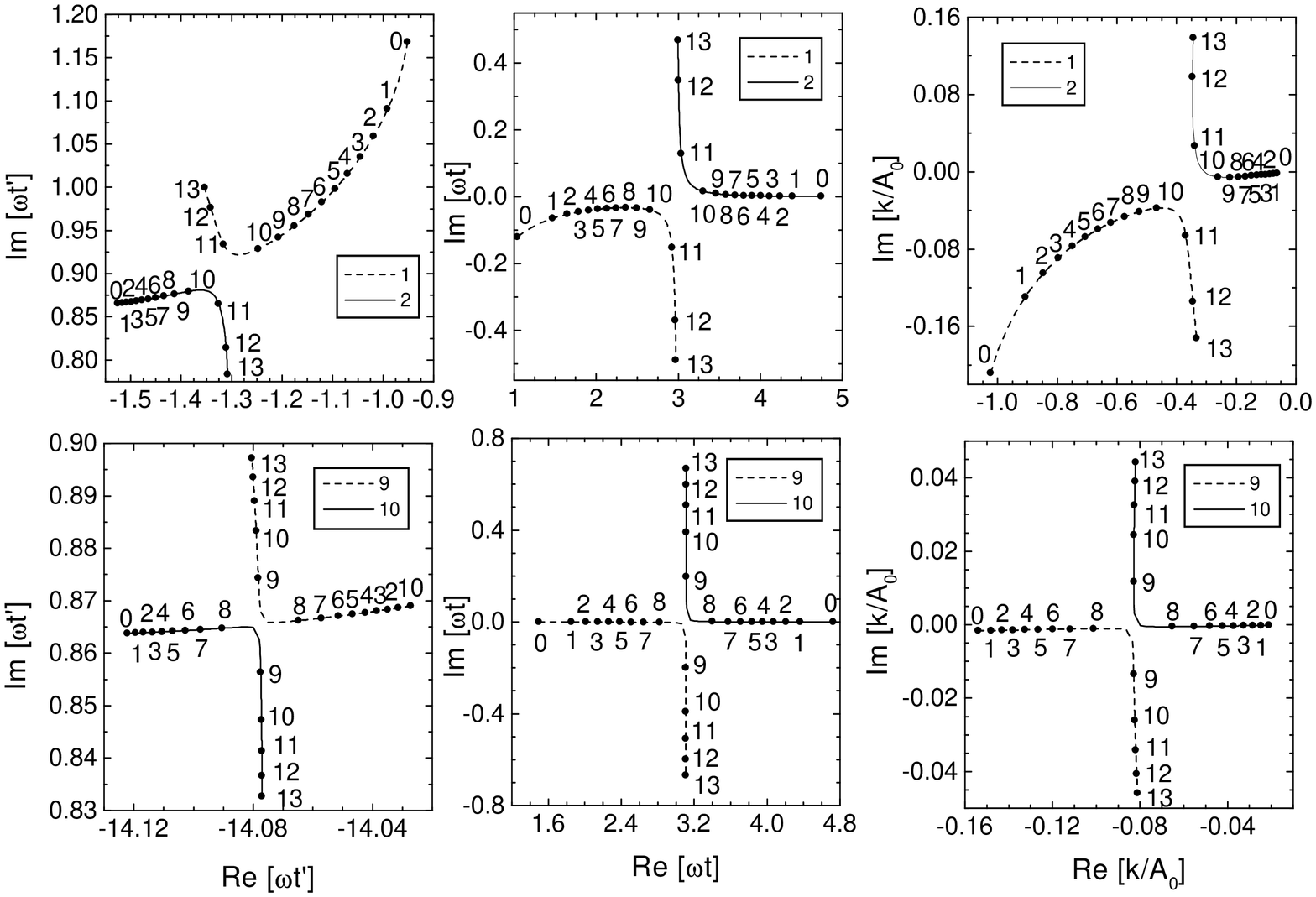,width=11.0cm,angle=270}
\caption{\label{figsaddles} Saddle points as a function of energy for a Keldysh
parameter of $\gamma=0.975$
and scattering angle $\theta=0$. The first, second and third column give the
start time, the return time, and the intermediate drift momentum, respectively.
The panels present the paths in the complex plane that are followed by the
saddle points as a function of the final energy, which is indicated by the
numbers, which are in multiples of $U_P$. The upper row gives the saddle points
for the pair of orbits with the shortest two travel times $(1+2)$, the lower
row for $(9+10)$, which is one of the pairs with the longest times considered 
in this paper.
 The figure shows how the saddle points of a pair approach each other
most closely near the classical cutoff. In each case, the contribution of the
orbit that is drawn dashed is dropped after the cutoff.}
\end{figure*}

Upon approach to the classical cutoff, the two solutions that make up one pair
come very close to each other. For an example, this is illustrated in Fig.
\ref{figsaddles}. The saddle-point approximation 
(\ref{sadresc}),
however, treats different saddle points as independent. As
mentioned in previous papers \cite{goresl}
and in the introduction,
this leads to a quantitative and qualitative
breakdown of the standard saddle-point approximation near the cutoff of any
pair of solutions, for two reasons:
(i) This approximation can overestimate the contribution to
the transition amplitude by several orders of magnitude (it actually
diverges if both saddles coalesce).
(ii)
In previous papers, the spurious
saddle has been dropped  after the classical cutoff 
by requiring a minimal discontinuity of the
transition amplitude. Still, the discontinuity remains
finite and noticeable.

A smooth suppression of the spurious saddle can be achieved
if both quantum orbits are well separated at the Stokes
transition (which is, however, not the case for physically accessible
parameters in ATI),
by a regularization that has been derived in the general framework of
asymptotic expansions
\cite{Berry}. Thereby, the contribution of the spurious saddle
 is suppressed by multiplication with the error function
\begin{equation}
\mathrm{erfc} (-\nu)=\frac{2}{\sqrt{\pi}}\int_{-\infty}^{\nu}d\tau
\exp (-\tau^2),
\label{eq:stokes}
\end{equation}
with the argument given by
\begin{equation}
\nu=\frac{\mathrm{Re}
\,[S_\vp(t_i,t^{\prime}_i)-S_\vp(t_j,t^{\prime}_j)]}{\sqrt{2|\mathrm{Im}\,
[S_\vp(t_i,t^{\prime}_i)-S_\vp(t_j,t^{\prime}_j)]|}}.
\end{equation}
The argument $\nu$ vanishes at the Stokes transition 
(\ref{stokesline}) and diverges at the
anti-Stokes transition (\ref{eq:antistokes}), 
after which the spurious saddle drops out completely.
Note that this  automatically prevents  an
exponential growth of the amplitude of the  spurious
saddle in the approximation (\ref{sadresc}), because
the saddle is dropped while the
imaginary part of the action is still positive (namely, equal to the imaginary
part of a physical saddle).

This regularization procedure is not accurate enough
in the present problem because the Stokes transitions take place while the
saddles are not sufficiently separated (cf.\ Sec.\ \ref{sec:compare}).
On the other hand,
the Stokes transitions are already built into the uniform approximation,
to which we turn now.

\section{The uniform approximation}
\label{uniform}

The saddle-point approximations (\ref{saddirect}) and
(\ref{sadresc})
are obtained by expanding
the action function $S_\vp$ to second order in the integration
variables about each saddle point, and then doing the ensuing Gaussian
integrals.
These approximations are valid if the expansion of the action
holds until the integrand has become much smaller than it was 
at the saddle point, so that the integration can be extended to infinity.
The saddle-point approximation breaks down 
when the difference of actions
$|S_i-S_j|$ of two quantum orbits with similar
coordinates becomes of order unity, such that the expansion about 
saddle point  $i$  becomes inaccurate close to the saddle point
$j$, and vice versa.
For the quantum orbits in ATI
this happens
when the energy approaches the classical cutoff.
The remedy offered by the theory of asymptotic expansions
is to improve the expansion of the action function
in the neighborhood of saddles $i$ and $j$
by including higher orders in the coordinate dependence
and to take the resulting approximate integral
as a collective contribution of both saddle points.

What is often not observed is that the resulting uniform approximation
can be written in such a form 
that no additional information on the
quantum orbits is needed, i.e., the cumbersome expansion
in the coordinate dependence actually can be circumvented.
The derivation proceeds in two steps. First, we write down the so-called
diffraction integral which describes a pair of orbits which might be
close to each other or well separated. Then, we determine
the  parameters of the formal expansion
in terms of  the quantities that enter the standard saddle-point approximation,
from the observation that
the conventional saddle-point approximation (\ref{sadresc})
has to be recovered in
the limit where the saddle points are sufficiently well separated.

For the first step,
we observe that it is precisely two quantum orbits that closely approach each 
other near each cutoff.
According to the splitting lemma of catastrophe theory \cite{poston}, the
parametrization of the integration domain can be rectified such that
the orbits approach each other along one of the (appropriately chosen) 
coordinate axes (denoted
by $x$ in the following).
This is the only direction where higher orders in the coordinate expansion
of the action have to be included, while the expansion in the other coordinates
can be restricted to second order such that these can be integrated
out by the usual saddle point
approximation [this is similar to integrating out $\vk$ in
the transition from Eq.\ (\ref{sadresc}) to Eq.\ (\ref{sadresc2})].
Hence the contribution of the pair of 
quantum orbits (denoted by $i$ and $j$) to the transition amplitude
can be reduced, in principle, to a one-dimensional
diffraction integral of the general form
\begin{equation}
M_{i+j}=\int_{c_l}^{c_u} dx\, g(x)\exp[i S(x)],
\label{eq:orig}
\end{equation}
where the action accounts for these two saddle points and the 
 integration boundaries
$c_u$, $c_l$ in (complex) infinity are assumed such that the
integrand decays to zero and the integral converges.
Moreover, an expression that reduces to 
the conventional saddle-point approximation when the quantum orbits
are well separated will be obtained if we allow for
a linear coordinate dependence in
the function $g(x)$. This motivates the use of the normal forms
(for a derivation in another semiclassical context, see Ref.\ \cite{Henning1})
\begin{equation}
S(x)=\bar S+\varepsilon x-a x^3,\qquad
g(x)=g_0+g_1 x.
\label{eq:expansion}
\end{equation}
Here we have chosen
the origin of the coordinate system exactly in the middle
between the two saddles, 
which have coordinates $x_{i,j}=\pm\sqrt{\varepsilon/3a}$
and coalesce when $\varepsilon=0$.

The uniform approximation that
we introduce here differs from an earlier
regularization method  \cite{goresl,GP00}, where the action was expanded
to cubic order about the stationary point corresponding to 
the classical cutoff. This led to the absence of the linear term 
in the function $g(x)$ in Eq.~({\ref{eq:expansion}). It is precisely this 
term whose presence allows us to match the standard saddle-point 
approximation both near  the cutoff and away from it.  
Thus, the method
of Ref.~\cite{goresl} coincides with the uniform approximation near the 
classical boundary, but deviates from it and from the exact solution 
farther away from
the cutoff region.

With expansion (\ref{eq:expansion}) inserted into the original integral
(\ref{eq:orig}), the amplitude $M_{i+j}$
reduces to a sum of Bessel functions,
\begin{eqnarray}
M_{i+j}&=& \sqrt{2 \pi \Delta S/3}\exp(i\bar S+i\pi/4)
\nonumber \\
&& {}\times
\left\{\bar A
[J_{1/3}(\Delta S)+J_{-1/3}(\Delta S)]
\right.
\nonumber\\
&&
\left.  {}+\Delta A
[J_{2/3}(\Delta S)-J_{-2/3}(\Delta S)]
\right\}
,
\nonumber
\\
\Delta S&=&(S_i-S_j)/2,\qquad \bar
S=(S_i+S_j)/2,
\nonumber
\\
\Delta A&=&(A_i-i A_j)/2,
\quad\bar A=(i A_i-A_j)/2,
\nonumber
\\
\label{eq:unif}
\end{eqnarray}
where the four independent parameters
$\bar S$,
$\Delta S=2 \varepsilon^{3/2}(27 a)^{-1/2}$, 
$\bar A=g_0(-2\pi i)^{1/2} a^{1/4} (3\varepsilon)^{-1/4}$, and
$\Delta A=g_1 (2\pi i)^{1/2}\varepsilon^{1/4} (3 a)^{-3/4}$
have been expressed by the amplitudes and actions that result from the
saddle-point approximation of the
diffraction integral (\ref{eq:orig}).

The uniform approximation is defined by inserting into Eq.\
(\ref{eq:unif}) the 
actions and amplitudes (\ref{sadresc}) of the respective pair of 
quantum orbits (which we denoted by $i$ and $j$).
We wish to stress that it is not necessary to
obtain the expansion parameters $\bar S$, $\varepsilon$, $a$, $g_0$, and
$g_1$
by explicitly carrying out the expansion (\ref{eq:expansion}).
Indeed, knowledge of the  explicit dependence on these parameters is not 
 even desired
because it can be manipulated by a coordinate transformation, while
the original integral
is invariant under smooth changes
of the coordinate system. 
For the saddle-point approximation (\ref{sadresc}),
invariance with respect to coordinate transformations is ensured 
trivially for the actions $S_s$,
while the amplitudes $A_s$ are invariant because the Jacobian of a 
transformation 
contributes a factor to $g$ which is
cancelled by the determinant of second derivatives of the action,
see Eqs.\
(\ref{sadrescc}) and 
(\ref{sadresc2c}).
This is the reason why we express
the expansion coefficients in Eq.\ (\ref{eq:expansion})
by the coordinate-transformation invariant quantities $A_{i,j}$,
$S_{i,j}$ of the saddle points.
Indeed, it is a simple exercise to verify
with the help of the asymptotic behavior
\begin{equation}
 J_{\pm \nu}(z)\sim \left(\frac{2}{\pi z}\right)^{1/2}
\cos(z \mp \nu\pi/2-\pi/4)
\end{equation}
of the Bessel functions for large $z$
that the saddle-point approximation
(\ref{sadresc}) is recovered from the uniform approximation
(\ref{eq:unif}) in the limit of large $\Delta S$.

Finally, let us demonstrate that the uniform approximation is also
capable of describing the Stokes transition, in which one of the two saddles
is rendered irrelevant.
The Bessel functions in (\ref{eq:unif})
assume complex arguments and are
multi-valued functions, depending on the integration contour taken in
their integral representation.
The functional branches can be
distinguished by the number of saddles which are visited by a
steepest-descent deformation of the contour,
in complete analogy with the procedure for the 
 original integral
(\ref{prescatt}).
Hence, when the condition (\ref{stokesline}) is fulfilled
one not only observes a Stokes transition in the original integral,
but also encounters a
Stokes transition in the defining integral of the Bessel functions.
The proper branch of the function 
will automatically be selected by requiring a smooth functional
behavior.
The choice of branches beyond the Stokes transition
corresponds to replacing the
Bessel $J$ functions by Bessel $K$ functions,
\begin{eqnarray}
M_{i+j}&=&\sqrt{ 2 i \Delta S/\pi}\exp(i \bar S)
\nonumber\\
&&{}\times \left[\bar A K_{1/3}(-i\Delta S)
+i\Delta A K_{2/3}(-i\Delta S)
\right].\quad
\label{eq:unif2}
\end{eqnarray}
 From the usual asymptotics
\begin{equation}
K_\nu(z)\sim \left(\frac{\pi}{2z}\right)^{1/2}\exp(-z)
\end{equation}
of the Bessel $K$ function
for large $z$ one verifies that in this case only saddle $i$ contributes to
the saddle-point approximation. 

In summary, in the uniform approximation the sum of
saddle-point amplitudes  (\ref{sadresc}) of each pair of quantum orbits
is simply replaced  by the collective amplitude (\ref{eq:unif}).
The uniform approximation improves the
saddle-point approximation such that it works even
when two quantum orbits approach each other so closely that one cannot
locally expand about either one, as is the case close to their
classical cutoff.
It also works well far away from classical
cutoffs, because it
includes the saddle-point approximation as
a special
case which is recovered for
$|\Delta S|\gtrsim 1$. This can happen in two ways:  (i) when the saddle points
become well separated as a system parameter (such as $\vp$) is varied, or (ii)
 in the strict
semiclassical limit when for fixed system parameter the Keldysh parameter is
decreased
(given $\Delta S \neq 0$).
Also, the Stokes transition at the classical cutoff is automatically
built into the uniform approximation.
Most notably, the uniform approximation is of the same practical simplicity
as the saddle-point approximation since it involves
the same amplitudes $A_s$ and actions $S_s$ defined in Eqs.~(\ref{sadresc}).

\section{Comparing the various approximations}
\label{sec:compare}

In this section, for the
zero-range potential we compare the approximations discussed in the previous
sections with the exact integration of Eq. (\ref{prescatt}).
 First, let us consider ATI spectra in the direction of the
electric field of the laser. Such a spectrum is composed of the contributions
of direct and of rescattered electrons. The former quickly decrease after their
classical cutoff at $2U_P$. The latter form an extended plateau with its
classical cutoff at $10U_P$, whose yield is below that of the direct electrons
by several orders of magnitude. The cutoff at $10U_P$ is related
to the pair of orbits with the shortest travel times. The other
pairs of trajectories, which have longer travel times,  have cutoff energies
below this value (see, e.g., Ref.~\cite{optcomm} for a more complete
discussion). In the figures
that follow, we consider up to 5 pairs of electron trajectories, those with the
shortest travel times. To each trajectory, we associate a positive integer 
number which increases with the corresponding travel time.

\begin{figure}[tbp]
\begin{center}
\epsfig{file=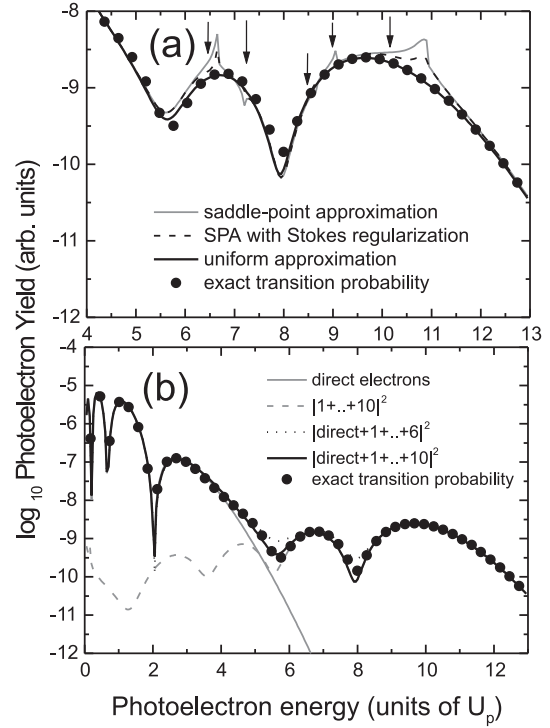,width=7.0cm,angle=0}
\end{center}
\caption{\label{comparison1}Photoelectron spectra
for a zero-range binding potential and $U_{P}/\omega =3.58,$ $
\omega =0.073\ {\rm a.u}.,$  and a ground-state energy of
$E_{0}=-0.5\ {\rm a.u}.$ The spectrum is in the direction of the electric field
of the laser, $\theta =0.$ Part (a) shows spectra computed using the
saddle-point and uniform approximations, compared with the
photoelectron yield obtained by computing the integral
(\protect{\ref{prescatt}})
exactly. We take into account the two direct trajectories and
five pairs of rescattered trajectories. The approximate energy positions
of the Stokes transitions, which coincide with
the respective classical cutoffs, are indicated by arrows. Part (b) displays
spectra computed by means of the uniform approximation, for direct,
rescattered, and both types of electrons, 
and compares these with the exact integration.}
\end{figure}

 The outcome of this comparison is displayed in
Fig.~\ref{comparison1}(a).
In general, there is a good
qualitative agreement between the saddle-point approximation
and the exact solution
(note, however, that the scale is logarithmic in this figure.)
Quantitatively, however, there are
marked discrepancies, which occur in those energy regions where the saddle
points that constitute a
particular pair approach each other and can no longer be treated as
independent.

In previous work \cite{optcomm}, the unphysical contribution of
one of the saddle points was eliminated by hand as soon as the energy crossed
the Stokes line (\ref{stokesline}).
This causes the cusps in the spectra, which can also be seen
in Fig. \ref{comparison1}(a).  
This is not very satisfactory, since the
discrepancies in the ATI signal may amount to almost one order of magnitude.
This problem is particularly critical if the intensity of the driving
field is not so high. In this case, the various cutoff energies  are
relatively close to each other, so that the artifacts affect  a
broad energy region.
Thus, a more accurate approximation is desirable and even necessary, in case
 the integral (\ref{prescatt}) cannot be carried out exactly, as is the case
for any potential
other than the zero-range potential.

One possibility to eliminate such effects, shown in
Fig.~\ref{comparison1}(a), is the 
Stokes regularization, Eq.\ (\ref{eq:stokes}).
This smoothes out the cusps, without, however,
eliminating them completely.

Far superior results are obtained by the uniform approximation,  
given by Eqs.\
(\ref{eq:unif}) and
(\ref{eq:unif2}),
respectively. The spectrum computed in this way almost perfectly agrees with
the
exact result.
The  remaining differences between 
the uniform approximation and
the exact integration occur near the interference minima and are due to
the contributions of pairs of trajectories with longer travel times that have
not been included. This is indicated by the
minor differences in the spectra computed with the uniform approximation
using 3 and 5 pairs of trajectories, cf. Fig. \ref{comparison1}(b).

Figure
\ref{comparison1}(b) shows that  the exact spectrum is well reproduced by the
uniform approximation for all energies.
The figure also separately displays the contribution of the direct
electrons \cite{footdir}. One observes that interference between the rescattered and
direct electron trajectories is only important within a  small energy region,
between $4U_{P}$ and $6U_{P}$ \cite{Petc00}. Above and below this energy range,
either the rescattered or the direct electrons completely dominate the
spectrum, so that interference only leads to minor effects.

\begin{figure}[tbp]
\begin{center}
\epsfig{file=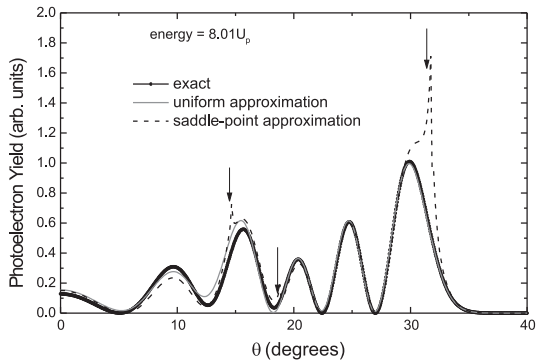,width=7.0cm,angle=0}
\end{center}
\caption{\label{comparison2}Angular distributions of photoelectrons for the
zero-range potential case, computed with the saddle-point and uniform
approximations, compared to the exact yield. The field parameters are
 $U_{P}/\omega =35.8$, $\omega =0.0584\ {\rm a.u}.$, the ground-state energy is
chosen as $E_{0}=-0.9\ {\rm a.u}.$, and the photoelectron energy is
$\epsilon=8.01 U_P$. The angles of Stokes transitions are
marked with arrows.}
\end{figure}

The superiority of the uniform approximation over
the saddle-point approximation becomes particularly impressive if spectra are
displayed on a linear scale. This is done in Fig.~\ref{comparison2} for an
angular distribution at fixed energy. Both with the saddle-point approximation
and the uniform approximation,
 the 10 shortest trajectories are considered. The uniform approximation,
again, yields excellent agreement with the exact
result. Minor differences,
 for small scattering angles, are caused by
the  trajectories with still longer travel times that have not been
included. Those  do not contribute for
larger angles.
The saddle-point approximation, on the other hand,
exhibits large discrepancies with the exact results near the classical
cutoffs.
For the chosen photoelectron energy of $8.01 U_P$, there are only three
relevant cutoffs,
corresponding to the pairs of trajectories 1+2, 5+6, and 9+10.
The remaining pairs of trajectories do not contribute, since their cutoffs 
are significantly below $8.01 U_P$.

\section{Influence of the potential on rescattering processes}
\label{sec:potential}

The preceding section has shown that the uniform approximation is a
very dependable method, yielding results very close to those obtained
from the exact integration. The latter, however, is only feasible
for a binding potential of zero range. Therefore, we will rely on the uniform
approximation to
investigate how the form of the binding potential affects
the photoelectron spectrum. The transition amplitude (\ref{rescatt}) 
was derived in the context of one electron bound by the potential $V(\vecr)$. 
In order to simulate a many-electron atom, it can be reasonable to use in 
the transition amplitude (\ref{rescatt}) different potentials $V(\vecr)$ 
for the electron when it tunnels out and when it rescatters \cite{GP00}. 
In Refs. \cite{GP00,goresl2}, the effect of the rescattering potential on the 
general shape of the high-order spectrum and the 
ratio of direct over rescattered electrons were investigated as a 
function of the applied field, for the pair of the two shortest orbits. In 
particular, the dependence on the atomic species was modeled by a Thomas-Fermi 
potential. Here, for various model potentials, making use of the 
additional power afforded by
 the uniform approximation, 
we will concentrate on the detailed shape of 
the angular-resolved energy spectrum and on the 
contributions of the orbits with longer travel times.
 
Throughout, we shall use the results for the zero-range
potential

\begin{equation}
V(r)=\frac{2\pi }{\sqrt{2|E_{0}|}}\delta (\vecr)\frac{\partial }{\partial
r}r \label{zerorange}
\end{equation}
as a benchmark. Its form factors (\ref{ffr}) and (\ref{ffi}) are constants,

\begin{equation}
V_{\vp \vk}=\frac{1}{(2\pi )^{2}\sqrt{2|E_{0}|}}  \label{vpkdelt}
\end{equation}
and
\begin{equation}
V_{\vk 0}=-\frac{(2|E_{0}|)^{1/4}}{2\pi }.  \label{vk0delt}
\end{equation}

\subsection{Influence of the Coulomb tail}
\label{coulombtail}
In this subsection, we investigate the influence of the long-range Coulomb
potential on
above-threshold ionization. This is particularly interesting since for
hydrogen ATI spectra have been extracted from  
a high-precision numerical solution of the
time-dependent Schr\"{o}dinger equation  (TDSE)
\cite{CormLam}, so that we can compare the strong-field approximation with an
exact solution.

The form factors of the Yukawa potential $V(r)=-Z \exp(-\alpha r)/r$ are

\begin{equation}
V_{\vp \vk}=-\frac{Z}{2\pi ^{2}}\frac{1}{(\vp-\vk)^{2}+\alpha ^{2}}
\label{formfyuk1}
\end{equation}
and
\begin{eqnarray}
V_{\vk 0}&=&-\frac{\sqrt{2}}{\pi }\frac{Z^{5/2}}{(Z+\alpha
)^{2}+[\vk+\vA(t^{\prime
})]^{2}} \nonumber\\
&=& -\frac{\sqrt{2}}{\pi }\frac{Z^{5/2}}{(Z+\alpha )^{2}-2|E_0|},
\label{formfyuk2}
\end{eqnarray}
where the saddle-point equation (\ref{saddle1}) has been used in the last line.
Hence, in the saddle-point approximation, $V_{\vk 0}$ acts as a constant;
indeed, this is the case for any spherically symmetric potential. This constant
determines the total ionization rate, but has no effect on the shape of the
spectrum. Another consequence is that the spectrum of the direct electrons,
described by the amplitude (\ref{pdirect}),
is independent of the form of the binding potential
because it only depends on $V_{\vk 0}$, in contrast to the
spectrum of the rescattered electrons.

The Coulomb form factors can be retrieved from Eqs.~(\ref{formfyuk1}) and
(\ref{formfyuk2}) in the limit $\alpha \rightarrow 0$. Since in this case
$E_0=-Z^2/2$, this leads to the well-known divergence of the Coulomb form
factor (\ref{formfyuk2}) \cite{lew94}. This has no effect on the shape of the
spectrum, and the \textit{absolute} scale can be reestablished, too \cite{rmp}.

\begin{figure}[tbp]
\begin{center}
\epsfig{file=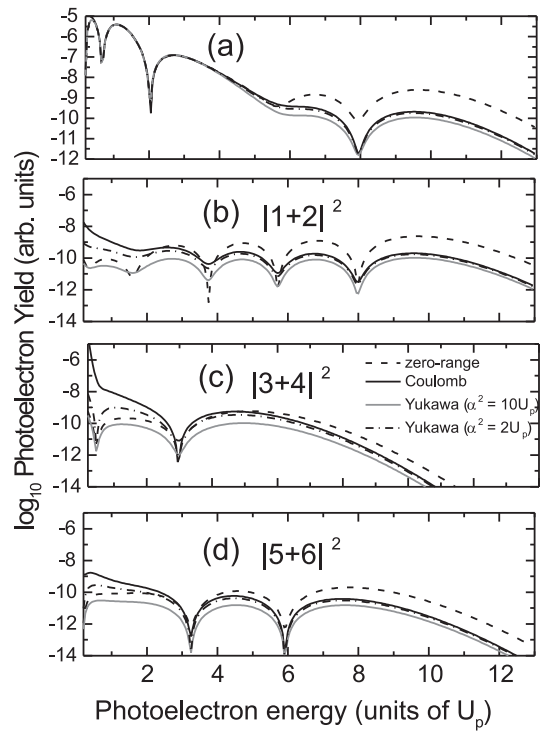,width=7.0cm,angle=0}
\end{center}
\caption{\label{potential1}Photoelectron spectra for the zero-range
potential, compared with those for  the
Coulomb and Yukawa potential, for the same field and atomic parameters as
in Fig.~\ref{comparison1}. Panel (a) shows total spectra, while panels (b) to
(d) exhibit the contributions of individual pairs of rescattered orbits.}
\end{figure}

\begin{figure}[tbp]
\begin{center}
\epsfig{file=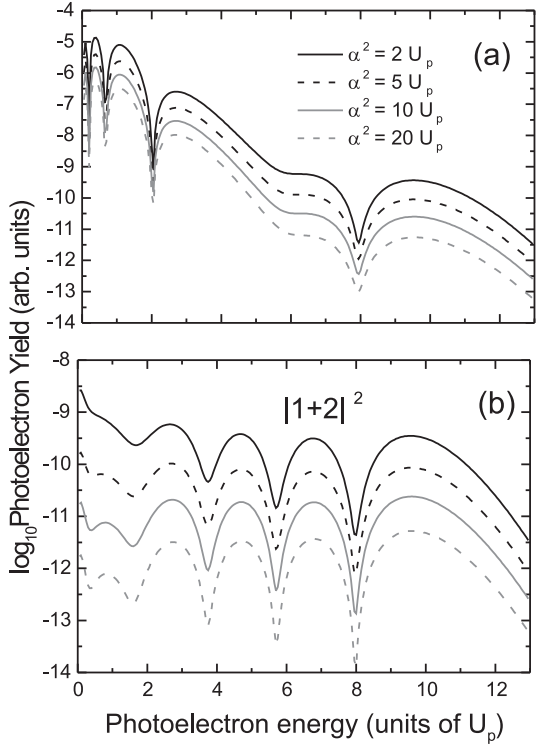,width=7.0cm,angle=0}
\end{center}
\caption{\label{potential2}Photoelectron spectra for the Yukawa potential,
the same field and
atomic parameters as in the previous figure, and several screening
parameters $\alpha$. Part (a) shows the resulting spectra for the direct 
electrons and the five shortest pairs of rescattered orbits,
whereas part (b) shows the contributions from the shortest pair of
rescattered trajectories.}
\end{figure}

 In Fig. \ref{potential1},
 we compare ATI spectra for the zero-range, the Yukawa, and
 the Coulomb potential. In view of the Coulomb divergence of
 $V_{\vk 0}$ we used the zero-range form factor (\ref{vk0delt}) for all
potentials \cite{footnote2}.
As expected from Eq. (\ref{formfyuk1}),
there is a suppression of the photoelectron yield for
the higher energies in the Coulomb and Yukawa cases. This effect is present
for all pairs of trajectories. For the Coulomb potential, there is an
additional enhancement of the rescattered yield for low energies, which does
not occur in the zero-range or short-range cases. This enhancement is due to
the functional form of $V_{\vp \vk}$. Clearly, if the screening parameter is
small enough, this effect is also present for the Yukawa potential.
Furthermore, for these latter potentials, there is a reduction in the plateau
intensity
as the screening parameter is increased. Evidently, the form factor
(\ref{formfyuk1}) for the Coulomb potential always exceeds the one for the
Yukawa potential.

The parameters of Figs.~\ref{comparison1} and \ref{potential1} correspond to those
chosen in Ref.~\cite{CormLam}, where the results of a  numerical solution of
the three-dimensional time-dependent Schr\"odinger equation for hydrogen are
reported and ATI spectra are extracted from the former. The agreement between the
Coulomb result of Fig.~\ref{comparison1} (a) and Fig. 2 of Ref.~\cite{CormLam}
is good and even quantitative. We notice that the pronounced dip in the
spectrum near $8U_P$, which is due to destructive interference of the
contributions of the shortest two orbits [cf. Fig.~\ref{potential1} (b)], is
almost at the same
position in both calculations. The next destructive-interference minimum from
these two orbits occurs just below $6U_P$. The contributions of the longer
orbits [cf. Fig.~\ref{potential1} (c) and (d)] partially fill in this minimum,
leaving only a shoulder in the total spectrum (a). The exact calculation
\cite{CormLam} features a slightly more pronounced minimum at the same
position.
Remarkably, the two interference minima in the total spectrum at low energy
near $0.5U_P$ and $2U_P$, which are due to the direct electrons and the
amplitude (\ref{pdirect}), are also clearly reflected in the exact calculation
\cite{CormLam} at about the same positions. The overall drop of the spectrum
from the direct electrons to the final maximum of the rescattered electrons
preceding the cutoff is more pronounced in the exact calculation by about 
half an order of magnitude \cite{anotherfootnote}.

In Fig.~\ref{potential2},
 we investigate the ATI spectra for several screening parameters
$\alpha $ of the Yukawa potential. In this figure, we also address
the question of how the form factor $V_{\vk 0}$ affects
the photoelectron yield.
The figure
clearly shows a global shift in the photoelectron signal, which increases for
decreasing $\alpha$. In this sense, our results are in agreement with those in
Ref.~\cite{milos}.
It is, however, not expected that this yield increases indefinitely. In fact,
its limit for $\alpha\to 0$ should be given by the TDSE results \cite{CormLam}.
Because of the singularity for hydrogen in $V_{\vk 0}$ for vanishing screening
parameters, such a comparison is beyond the scope of the strong-field
approximation.
 Additionally, there is an enhancement
of the photoelectron yield at lower energies, similar to those
occurring in the
Coulomb case, which disappears as $\alpha$ is increased, which is in agreement
with the previous figure.

\subsection{Shell potentials}

Spherical shell potentials have been  
used for modeling clusters or
molecules such as C$_{60}$. Recently,
ATI  has been observed experimentally for C$_{60}$ in
the direct-electron 
energy region \cite{ATIclust}.
Therefore, in this section we investigate how such potentials affect the ATI spectra 
in the
direct and rescattered regions. Let us first consider a spherical $\delta$-shell,
\begin{equation}
V(r)=-V_{0}\delta (r-r_{0}),  \label{delts}
\end{equation}
with
\begin{equation}
V_{0}=\frac{\sqrt{2|E_{0}|}}{1-\exp [-2\sqrt{2|E_{0}|}r_{0}]},
\end{equation}
where $E_0$ again denotes the binding energy of the ground state. Ionization
from such a potential was investigated in the past \cite
{deltashell}, for weaker laser fields.
The corresponding  form factors (\ref{ffr}) and (\ref{ffi})
are
\begin{equation}
V_{\vp\vk}=-\frac{V_{0}r_0}{2\pi^2\sqrt{(\vp-\vk)^2}}\sin
[\sqrt{(\vp-\vk)^2}r_{0}] 
\label{vpkdeltas}
\end{equation}
and
\begin{equation}
V_{\vk 0}=-\frac{V_{0}C}{\pi  \sqrt{|E_0|}r_{0}}
\sinh (\sqrt{2|E_0|}r_{0}),
\label{vp0deltas}
\end{equation}
respectively, with
\begin{equation}
C=\left[\frac{\sqrt{2|E_0|}}{\exp(2\sqrt{2|E_0|}r_0)
-1-2\sqrt{2|E_0|}r_0}\right]^{1/2}.
\end{equation}
For the $\delta$-shell potential, $V_{\vp\vk}$
is an oscillating
function, and $V_{\vk 0}$ is a constant as always. 
Thus, in the following, we concentrate on the influence of
 $V_{\vp\vk}$
on the resulting spectra. We consider typical C$_{60}$
parameters, taken from Ref. \cite{ATIclust}. The external field is chosen
such that its intensity is still
below the C$_{60}$ fragmentation threshold, but the electron excursion
amplitude \cite{footnexcursion} is roughly twice as large as $r_{0}$.
Furthermore, the Keldysh parameter
is about unity. Thus, the rescattering picture is still expected to be
applicable.
\begin{figure}[tbp]

\begin{center}
\epsfig{file=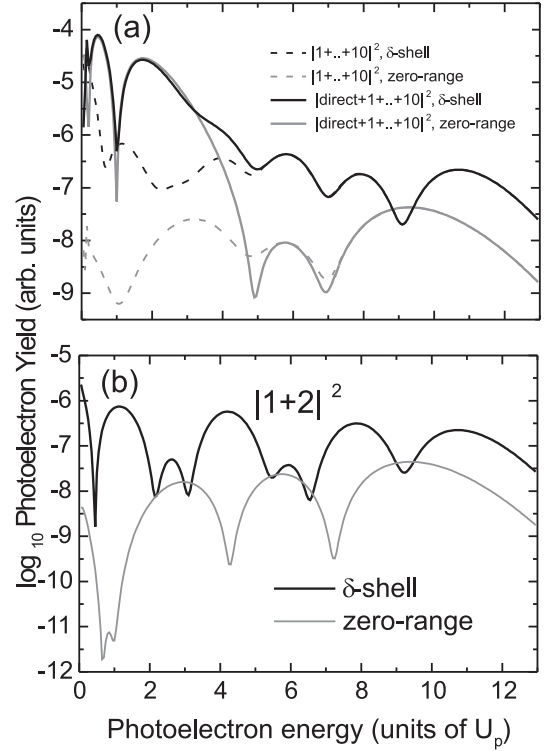,width=7.0cm,angle=0}
\end{center}
\caption{\label{cluster1}Photoelectron spectra for the shell potential
(\protect{\ref{delts}}), compared with the zero-range case. The ionization
potential was taken as $|E_0|=0.274$ a.u. and the cluster radius as $r_0=6.7$
a.u. The field parameters are $I_0=6.5 \times 10^{13}\mathrm{W/cm^2}$, and
$\omega=0.057$ a.u. This yields an excursion amplitude of $a_0=13.2$ a.u. and a
Keldysh parameter $\gamma=0.9805$. In part (a) we
take into account the five shortest pairs of trajectories, whereas in part
(b) only the shortest pair is considered.}
\end{figure}

In Fig.~\ref{cluster1}, we compare the photoelectron spectrum for the $\delta
$-shell and for the zero-range potential,
within the uniform approximation.  In order to assess the efficiency of
rescattering, in either case we used for $V_{\vk 0}$ the zero-range result
(\ref{vk0delt}). The figure shows that the $\delta$-shell potential rescatters
more efficiently than the zero-range potential by about one order of magnitude.
If the form factor (\ref{vp0deltas}) is taken into account, an additional
global increase in the yield occurs. However, in the $\delta$-shell case, the
rescattering plateau on the average has a downward slope, in contrast to the
zero-range case where the slope goes up.

The most interesting feature,
however, is that
the rescattered spectrum of the $\delta $-shell potential is much more
structured than it is for the zero-range potential, with several additional
oscillations. Such oscillations are due to the form factor (\ref{vpkdeltas}),
and are already present for the contributions of the shortest pair of
trajectories, as shown in Fig. \ref{cluster1}(b). An unexpected side effect of
these oscillations is the effective increase of the plateau cutoff energy by
about two units of $U_P$ for the shell versus the zero-range potential,
which can be observed in Fig.~\ref{cluster1}. Since the laser intensity is the
same in both cases, the rescattering cutoff would be expected at the same
energy, too. However, the shell form factor has a zero around the energy of
$9.5U_P$, where the zero-range spectrum features its final maximum. This moves
the final maximum of the shell-potential spectrum up to a higher energy.

In order to investigate these oscillations in more detail, in the following we
will look at contributions of \textit{individual}
trajectories to the photoelectron yield for the $\delta $-shell, in
comparison to the zero-range potential. Since the uniform approximation
requires pairs of trajectories, we will use the saddle-point approximation for
that
purpose.
 Whenever dealing with a {\it pair} of
trajectories, we will consider the uniform approximation.

\begin{figure}[tbp]
\begin{center}
\epsfig{file=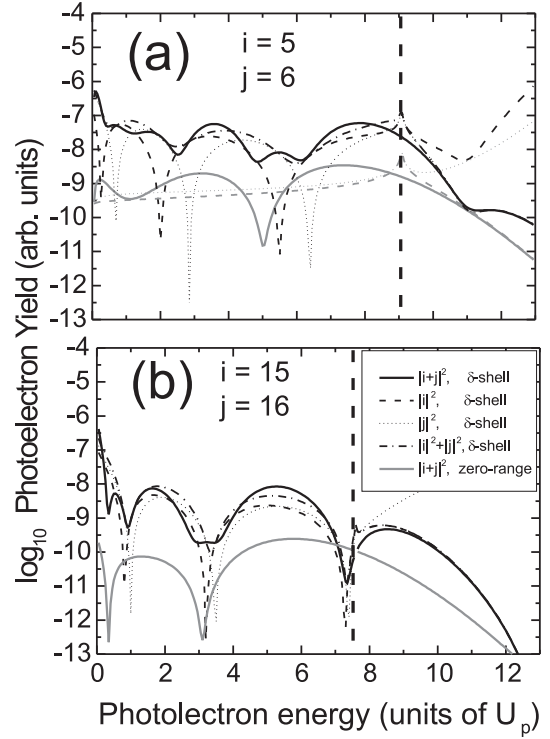,width=7.0cm,angle=0}
\end{center}
\caption{\label{cluster2}Contribution from individual trajectories to the
rescattered
photoelectron spectrum for the shell potential, in comparison to the zero-range
case. We consider the
same parameters as in the previous figure. The labels $i$ and $j$ refer, in
part (a),  to the third-shortest pair, denoted by (5+6), and in part (b) to the
eighth-shortest pair,
denoted by (15+16).
For the terms $|i+j|^2$ we applied the uniform approximation.
For the terms $|i|^2+|j|^2$, we applied the Stokes
regularization (\ref{eq:stokes}) to the
diverging trajectory. The dashed vertical lines in the figure separate the
classically allowed and forbidden energy regions for the respective orbits. The
dotted and dashed gray lines in part (a) denote the individual contributions of
5 and 6, respectively, for the zero-range case.}
\end{figure}

Figure \ref{cluster2} displays these results, for several rescattered
trajectories. In
case $V_{\vp \vk}$ is constant, as is the case for the zero-range
potential, all oscillations present in the spectra come from interference
terms. The contributions of individual trajectories are nearly constant 
in the classically 
allowed regime and do not produce any  substructure.
 For the $\delta$-shell, however, $V_{\vp \vk}$ is
oscillatory and produces its own
maxima and minima in the spectrum. However, comparing Fig.~\ref{cluster1}(a)
and (b) we observe that the contributions of the longer orbits tend to restore
the minima of the shell-potential spectrum to those of the zero-range. Only the
highest-energy minimum near $9.5 U_P$ is left unaffected, since the longer
orbits do not contribute to this energy.

In particular, the minima are given by $\textrm{Re}\, \sqrt{(\vp-\vk)^2}=n\pi
/r_{0}$, where
$n$ is an integer. To a first approximation, the drift momentum $\vk$ can be
neglected with respect to the momentum $\vp$, so that the
energy positions of the minima, in units of the ponderomotive energy, are
roughly given by
\begin{equation}
\frac{p^{2}}{2U_{P}}=\frac{n^{2}\pi ^{2}}{r_{0}^{2}U_{P}}.  \label{rough1}
\end{equation}
This expression is expected to work better for longer excursion times, since,
according to Eq.~(\ref{saddle3}),
$\vk\propto 1/(t-t^{\prime })$ .
This can already be seen in Fig.~\ref{figsaddles}, where the saddle points
as functions of the energy are depicted. For a pair of trajectories with short
travel times,
the start and the return times, as well as the intermediate momentum {\bf k},
vary considerably with the photoelectron energy. For a long travel time, on the
other hand,
these quantities are nearly constant, in the classically allowed region.
Furthermore,
the return time, as well as the intermediate momentum, are almost real and
$\vk$ is very small.

Clearly, there exist deviations
from Eq.~(\ref{rough1}) due to the fact that $\vk$ is non-vanishing and
complex, $t$ and $
t^{\prime }$ are complex, and due to the time dependence of the
intermediate momentum. For instance, a feature that is not explained by
Eq.~(\ref{rough1}) is a shift in the oscillations of the longer
trajectory, with respect to those of the short one. This feature occurs for
all pairs of trajectories, and decreases as the travel times get longer.
A qualitative estimate of these deviations can be obtained by considering
$\sqrt{(\vp-\vk)^2}$ up to first order in $\sqrt{{\bf k}^2}$, and the pair  $
(t_{1},t_{1}^{\prime })$ and $(t_{2},t_{2}^{\prime })=(t_{1}-\varepsilon
,t_{1}^{\prime }+\varepsilon ^{\prime })$ up to first order in $\varepsilon
,\varepsilon ^{\prime}.$ This gives a shift in the minima, which is
proportional to $\varepsilon /(t-t^{\prime }),$ confirming the results
presented in Fig.\ \ref{cluster2}.

Now we turn to other shell potentials.
Similar results are obtained for a more realistic
square well, of the form $V(r)= -V_0$ for $r_{1}<r<r_{2}$, and zero otherwise.
Since, in nature, the sharp edges present for a $\delta$-shell or a square
well are smoothed out, it is of interest to investigate whether the
additional oscillations are also present for smooth potentials that
approximate Eq.~(\ref{delts}). One such example is the Gaussian potential
\begin{equation}
V(r)=-V_{0}\exp [-(r-r_{0})^{2}/\sigma^2 ].
\label{gaussian}
\end{equation}
For vanishing width, we recover Eq.~(\ref{delts}). For this potential, the
form factor $V_{\vp\vk}$ is given by a rather complicated expression,
which will
not be reproduced here. Important features of $V_{\vp\vk}$
are the presence of minima and a decrease with increasing
asymptotic momentum. This
 decrease dampens the oscillations, such
that $V_{\vp\vk}$, in comparison to the $\delta$-shell form factor, decays
much more rapidly for large $\vp$. This effect becomes more pronounced
as the width of the potential increases.

\begin{figure}[tbp]
\begin{center}
\epsfig{file=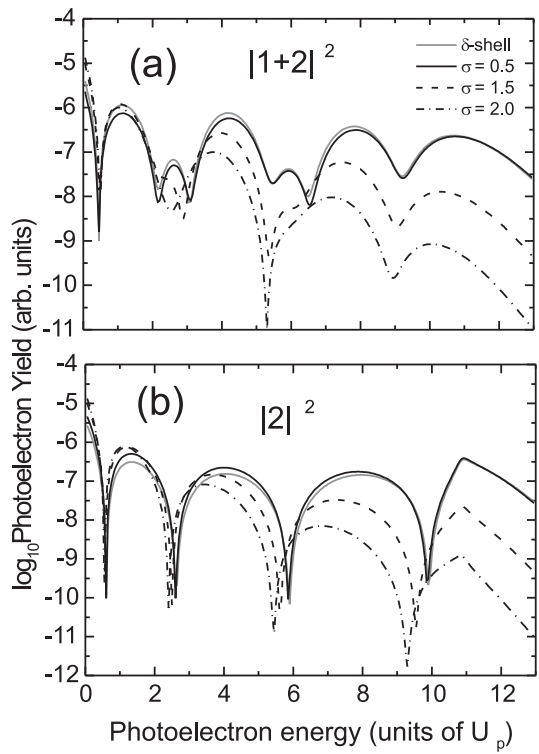,width=7.0cm,angle=0}
\end{center}
\caption{\label{cluster3}
Contribution from the shortest pair of trajectories to the
photoelectron spectrum for the Gaussian potential (\protect{\ref{gaussian}}),
compared to the $\delta$-shell case,
for several widths $\sigma$ and the same parameters as in the
previous figure. Parts (a) and (b) depict $|1+2|^2$ and $|2|^2$, respectively.
The prefactor $V_0$, for $|E_0|=0.274$, was computed by solving
the time-independent Schr\"odinger equation numerically.}
\end{figure}

In Fig. \ref{cluster3}(a), the contribution of the two shortest
trajectories to the ATI spectra is displayed for the Gaussian potential
(\ref{gaussian}), in comparison to the $\delta$-shell
potential.
We considered the zero-range-potential form factor $V_{\vk 0}$ [Eq.
(\ref{vk0delt})]. As in the previous figure, there exist additional
oscillations, which come from $V_{\vp \vk}$. In Fig. \ref{cluster3}(b),
this is clearly shown, for the contributions from the second shortest
trajectory.
For small width, as expected, the $\delta$-shell oscillation pattern is
practically recovered. For the parameter range considered in the figure,
this holds for $\sigma\alt 0.5$. Major differences are present only for
$\sigma > 1.5$. As the width gets larger,
there is a displacement in the minima
of the form factor and a suppression of the photoelectron yield.
This suppression is due to the decay of the form factor $V_{\vp\vk}$.
Therefore, even when the shell potentials are smoothed out, the
oscillations survive. Thus, the possibility that they are artificially
caused by the sharp edges of the $\delta$-shell potential can be ruled out.

\section{Conclusions}
\label{sec:conclusions}

We investigate the influence of the binding potential in above-threshold
ionization
(ATI) for linearly polarized laser fields, in terms of quantum orbits,
 using the  uniform
approximation [Eqs. (\ref{eq:unif}) and (\ref{eq:unif2})]. In this method, 
the transition amplitude is expanded in terms of the collective contribution of
pairs of orbits rather than individual orbits.
No information is required beyond  the
conventional saddle-point approximation. This is made possible and, indeed,
necessitated by the fact that for laser-induced rescattering
phenomena the orbits naturally come in pairs that nearly coalesce at the classical cutoffs, thus rendering the conventional saddle-point approximation inapplicable in this energy region.
Moreover, the uniform approximation 
remains valid beyond the
classical cutoff in the classically forbidden region, where it automatically incorporates the 
fading out of unphysical saddles beyond the cutoff energy. If the two saddles of a pair are sufficiently far apart, the standard saddle-point approximation is recovered.

The fact that the uniform approximation is valid in the whole energy
range, both away  from as well as near the cutoffs, allows one to obtain
quantitative predictions for ATI spectra. Indeed,
in this paper this approximation
has been tested for the zero-range potential against the
numerical computation of the SFA transition amplitudes. 
The photoelectron spectra,
as well as the angular distributions obtained in both ways turned out to be
practically identical.
With the conventional saddle-point approximation,
quantitative predictions are not possible in certain energy regions, which
for low laser intensity can span the better part of the ATI plateau.

The excellent quality of the uniform approximation 
for the zero-range potential
also suggests that the uniform approximation is reliable
enough for computing ATI spectra for other binding potentials, such as
Coulomb, Yukawa, or shell potentials.
Within the framework of this paper, the influence of the binding potential
is contained in two form factors, which
either characterize the transition from the ground state to an intermediate
momentum state, or the
transition from the intermediate state to an asymptotic momentum state.
Throughout the paper, these
form factors are called $V_{\vk 0}$ and $V_{\vp \vk}$, respectively.

As a first application, we investigated the role of the Coulomb tail by
computing photoelectron spectra for Coulomb and Yukawa potentials. As
a main feature, we observe a suppression of the photoelectron yield for the
ATI plateau, in comparison to the zero-range case, for both Yukawa and
Coulomb cases. This is due to the functional forms of $V_{\vp \vk}$, which
 are inversely proportional to the photoelectron momentum.
Additionally,  for the Coulomb potential
this form factor causes an increase in the low-energy ATI peaks.
These results are in agreement with the
fully numerical solution of the time-dependent Schr\"odinger equation
\cite{CormLam}. Furthermore, for the Yukawa potentials,
we observed an increase in the yield for decreasing screening parameter.
Similar features have been obtained in \cite{milos}, from the numerical
solution of the strong-field approximation transition amplitudes.

Another class of potentials that we investigated are shell potentials,
which are commonly used as an approximation for clusters. In comparison
to the zero-range case, the photoelectron spectra computed for such
potentials exhibit additional structure, which comes from the oscillating
form of $V_{\vp \vk}$. This is an extreme case of how the form factor
$V_{\vp \vk}$ influences the photoelectron yield. Such oscillations are
also present when the potentials are smoothed out, and therefore
are not an artifact of the shell models.

 An alternative for performing such
investigations is the numerical solution of the three-dimensional
Schr\"odinger equation. This would require considerable numerical effort, and,
for elliptical polarization, it would take one close to the limit of today's
computational resources. Another possibility would be the numerical solution
of the strong-field approximation amplitudes  (\ref{direct}) and
(\ref{rescatt}). From the
numerical viewpoint, this is not an easy task either, since one must deal with
multiple integrals of highly oscillating functions. Thus, the uniform
approximation considerably simplifies the computations involved.
Furthermore, using this approximation, one is able to gain
additional physical insight into the interference processes between the
quantum orbits, and how such processes are affected by the
binding potential.

Summarizing, the uniform approximation is a very powerful method
for investigating laser-assisted rescattering processes, being applicable
in all energy regions of the spectra.
This approximation allows one to compute photoelectron spectra for 
binding
potentials other than the zero-range with minimal numerical effort. 
Application of the 
methods developed in this paper to other high-intensity
laser-induced or laser-assisted phenomena, such as non-sequential double ionization, or to elliptically polarized fields is, in principle, straightforward.
\begin{acknowledgments}
This work was supported in part by the Deutsche Forschungsgemeinschaft. We are
grateful to S. P. Goreslavskii and S. V. Popruzhenko for useful discussions, 
to S. V. Popruzhenko for the critical reading of the manuscript, to 
M. E. Madjet for
providing
references on clusters, to R. Kopold for giving us his code for computing
the exact results, and to A. N. Salgueiro for her collaboration in the early 
stage of this project.
\end{acknowledgments}


\begin{thebibliography}{99}
\bibitem{footnkeldysh}In practice, the tunneling picture is still applicable if $\gamma \alt 1$.
\bibitem{footnote27} Both imaginary parts are positive:  As a rule, any 
saddle that has a negative imaginary part of the action
 cannot be visited by the steepest-descent contour.
 This is so because in the original integration the action
 was real $(\mathrm{Im}\, S=0)$, and the deformation procedure
does not lead to an increase of $\mathrm{Im}\, S$, by construction.
\bibitem{footnote2} Note that, in addition, for the same parameters as
in  \cite{CormLam}, the matrix element (\ref{formfyuk2})
is singular, such that a direct comparison would not be possible.
\bibitem{anotherfootnote} Note that our results include the phase-space factor;
that is, all of our spectra plot the quantity $|\vp|\, |M|^2$.
\bibitem{footnexcursion}The excursion amplitude $a_0=A_0/\omega$ is the
amplitude of the oscillatory motion of a classical electron in the laser 
field (\ref{eq:linpol}).
\end{thebibliography}
\end{document}